\documentclass[11pt]{article}

\usepackage{fullpage}               
\usepackage{graphicx,epsf,subfigure}
\usepackage{pstricks,pst-node,psfrag}
\usepackage{algorithm,algorithmic}
\usepackage{setspace}
\usepackage{amsfonts}
\usepackage{amsthm,amssymb,amsmath}
\usepackage{bm,url}
\usepackage{threeparttable,booktabs}
\usepackage{verbatim,url}
\usepackage{multirow}
\usepackage{fancyhdr}
\usepackage{natbib}
\usepackage{soul}

\DeclareMathOperator*{\argmax}{arg\,max}

\begin{document}

\begin{flushright}
Article (Methods)\\
\end{flushright}

\begin{center}
\begin{Large}
{\bf A phylogenetic approach disentangles interlocus gene conversion tract length and initiation rate}
\end{Large}
\end{center}

\begin{center}
Xiang Ji$^{\ast, 1,2,3}$, Jeffrey L. Thorne$^{\ast, 1,2,4}$ \\

\small $^{1}$Bioinformatics Research Center, North Carolina State University, Raleigh, NC, 27695\\
\small $^{2}$Department of Statistics, North Carolina State University, Raleigh, NC, 27695\\
\small $^{3}$Current Address: Department of Biomathematics, University of California, Los Angeles, CA, 90095\\
\small $^{4}$Department of Biological Sciences, North Carolina State University, Raleigh, NC, 27695\\
\small $^{*}$Correspondence: xji3@ucla.edu, thorne@statgen.ncsu.edu
\end{center}

\date{}


\begin{abstract}
Interlocus gene conversion (IGC) homogenizes paralogs.  Little is known
regarding the mutation events that cause IGC and even less is known about the
IGC mutations that experience fixation.
To disentangle the rates of fixed IGC mutations from the tract lengths of these
fixed mutations, we employ a composite likelihood procedure.  We characterize
the procedure with simulations.  We apply the procedure
to duplicated primate
introns and to protein-coding paralogs from both yeast and primates.
Our estimates from protein-coding data concerning
the mean length of fixed IGC tracts were unexpectedly low and are
associated with high degrees
of uncertainty.
In contrast, our estimates from the primate intron data had lengths in
the general range expected from IGC mutation studies.
While it is challenging to separate the rate at which fixed IGC mutations
initiate from the average number of nucleotide positions that these
IGC events affect,
all of our analyses indicate that IGC is responsible for a
substantial proportion of evolutionary change in duplicated
regions.  Our results suggest that
IGC should be considered whenever the evolution of multigene families is
examined.


\end{abstract}


\newpage

\section{{Introduction}\label{sec:Intro}}

Interlocus gene conversion (IGC) homogenizes repeats by
copying a tract of sequence from one paralog
to the equivalent region of another.
This means that evidence of nucleotide substitution in one paralog can be
erased and that the ancestry of an IGC tract coalesces in two paralogs
when IGC events occur. As a result, IGC events partition the sequence
sites of a multigene family into regions that have different evolutionary trees.
This consequence of IGC has complicated the study of multigene family evolution,
especially when the goals are to examine orthology
and the history of gene duplication and loss.
Incorporating tract length into IGC inference can therefore be helpful
for disentangling the local correlation structure of the histories
of sequence sites.

In \cite{Ji2016}, we introduced an approach to incorporate IGC into
any existing nucleotide substitution model by jointly considering
the corresponding nucleotide or codon sites in different paralogs.
One limitation of this modeling framework is that it assumes IGC events are
independently experienced by sites within a paralog.
The framework does not reflect the correlation structure among sites
within a paralog that is induced by IGC and we will refer to
it as the independent-site (IS) approach.
Here,
we extend the IS model by incorporating
IGC tract information and introduce an accompanying inference procedure.
We then illustrate the extensions
of our IS approach by applying them to three diverse groups of data sets
and by analyzing simulated data.

\section{{New Approach} \label{sec:NewApproach}}

The IS model can incorporate IGC by adding one new parameter ($\tau$) to
any conventional substitution model that has changes originate with
point mutation.  The purpose of the additional parameter is to represent
the homogenization among paralogs caused by IGC.
For example,
the HKY model \citep{Hasegawa1985} describes nucleotide substitutions that originate
with point mutation and has substitution rates
depend on the type of nucleotide being introduced and whether the
substitution is a transition or a transversion.
The HKY rate from nucleotide type $i$ to type $i'$ ($i \neq i'$)
is $Q_{ii'}$ with
\begin{equation}
{Q_{ii'}} \propto  \left\{ {\begin{array}{*{20}{c}}
{{\pi _{i'}}}&{{\mbox{if transversion}}}\\
{\kappa {\pi _{i'}}}&{{\mbox{if transition}}},
\end{array}} \right.
\label{Eq:HKY}
\end{equation}
where $\pi_{i'}$ is the stationary probability of nucleotide type $i'$
($\pi_A+\pi_C+\pi_G+\pi_T =1$) and $\kappa \geq 0$ differentiates
transitions and transversions.
With two paralogs per genome, the IS extension of the HKY model has
$Q_{(i,j),(i',j')}$ be the instantaneous rate at which one paralog
changes from state $i$ to $i'$ and the corresponding site of the other paralog
changes from state $j$ to $j'$.  The resulting rates for possible changes are:
\begin{equation}
{Q_{(i,j),(i',j')}} \propto  \left\{ {\begin{array}{*{20}{c}}
0&{i \neq i',j \neq j'}\\
Q_{ii'} & {i \neq i', j=j', i' \neq j'}\\
Q_{ii'}+ \tau & {i \neq i', j=j', i' = j'}\\
Q_{jj'} & {i = i', j \neq j', i' \neq j'}\\
Q_{jj'} + \tau & {i = i', j \neq j', i'=j'}.
\end{array}} \right.
\label{Eq:HKY+IS-IGC}
\end{equation}
IGC extensions to other conventional nucleotide or codon substitution models can be made in a similar fashion.

While the IS approach
considers dependence due
to IGC at corresponding sites in different paralogs, it
assumes IGC occurs independently at different sites within
the same
paralog. As \cite{Nasrallah2010} noted,
the assumption of evolutionary independence among sites
can be biologically unreasonable,
but is often kept for simplicity and computational
tractability.
Ideally, dependent evolution among sequence positions would be modeled
by treating entire sequences as the state of a
system (e.g., \citealt{Robinson2003}).
Having the state space consist of all possible sequences would
lead to more realistic Markov models for describing
how individual IGC events can affect multiple sites within a paralog.
However, the size of the state space becomes large when it matches
the number of possible sequences. This large size would not prove
computationally tractable if employing the inference strategy
outlined in \cite{Ji2016}.

As a compromise between computational feasibility and realism,
our new approach extends the IS model by jointly
considering pairs of sites in a paralog as well as
the corresponding pairs of sites in each other paralog.
We refer to the new approach for separately estimating
both the distribution of IGC tract lengths
and the IGC initiation
rate
as the pair-site (PS) approach.
The PS approach is a composite likelihood procedure that
statistically resembles the
population genetic technique of \cite{McVean2002} for estimating homologous
recombination rates.

\subsection{Relaxation of the site-independent assumption} \label{sec:TractModelSetup}

The parameter $\tau$ of the IS approach (e.g.,
see Equation~\ref{Eq:HKY+IS-IGC}) can be interpreted
as the average rate at which a site experiences IGC.
This average can be further decomposed into two factors.
The first is the rate per site of initiation of
IGC events that are destined for fixation. The second
is the average length of an IGC tract that becomes fixed.
We note that the average length of a fixed IGC tract may be
less than the average tract length of an IGC mutation because,
subsequent to an IGC event, the sequence may experience homologous
recombinations that cause the tract lengths of IGC mutations to
differ from tract lengths of fixed IGC events.
Here, we concentrate on inference regarding fixed IGC tracts.
We ignore the possibility of fixation of non-contiguous sequence
stretches arising from the tract produced by a single IGC mutation.

We model the rate per site at which fixed IGC events
initiate as being $\eta$.
We use a geometric distribution with parameter $p$ to model
the fixed IGC tract length distribution so that the probability
of a fixed IGC event covering $k$ sites is $p{(1 - p)^{k - 1}}$  with $k$
being a positive integer.
The mean fixed tract length is therefore $\frac{1}{p}$.
Although IGC events cannot initiate at a position that is $5'$
of a duplicated region and cannot extend $3'$ of a duplicated region,
we ignore these ``edge effects'' by having the expected IGC rate
be identical at all sites. This treatment can be interpreted as having
some IGC tracts initiate $5'$ of the sequence regions being followed and/or
terminate $3'$ of those regions.
With this treatment,
the rate at which a site experiences IGC can be written as $\tau  = {\eta }/{p}$.
The IS model can be viewed as the special case
where $p = 1$.

\subsection{The PS approach} \label{sec:PS-IGC}

Because computational constraints hinder the ability to employ
codon-based substitution models, we implemented our PS expansion in
conjunction with a modified HKY model \citep{Hasegawa1985}
where the HKY rates are
employed to describe nucleotide substitutions that
originate with point mutation.
Other conventional nucleotide substitution models
could be used instead.
\cite{Harpak2017} have independently developed a composite likelihood procedure
that extends the IS model of \cite{Ji2016} to IGC tracts.  Their approach
is very similar to the one here (see also \citealt{Ji2017}),
but \cite{Harpak2017} achieve computational feasibility by reducing
the $4$ nucleotide types to binary characters.  This state space reduction
may be especially problematic when sequences being analyzed
are not closely related.

To better adapt our modified HKY model
to analysis of protein-coding sequences, we add parameters that permit
rate heterogeneity among codon positions.  Specifically, the parameters
denoted $r_2$ and $r_3$ respectively
represent the ratios of fixation probabilities at second
and third codon positions relative to the first codon position.
The resulting model has rates $Q_{ii'}$ of fixed point mutations from
nucleotide type $i$ to $i'$ being
\begin{equation}
{Q_{ii'}} \propto  \left\{ {\begin{array}{*{20}{c}}
{r{\pi _{i'}}}&{{\mbox{if transversion}}}\\
{r\kappa {\pi _{i'}}}&{{\mbox{if transition}}},
\end{array}} \right.
\label{Eq:ModifiedHKY}
\end{equation}
where $r = 1$ for the first codon position,
$r = r_2$ for the second codon position
and $r = r_3$ for the third codon position.
We will refer to this as the independently-evolving paralog model (HKY-IND)
in order to contrast it with our models
that add dependence among paralogs
due to IGC.
The IGC treatment
that combines the IS parameterization of Equation \ref{Eq:HKY+IS-IGC}
with the
$Q_{ii'}$ rates of Equation \ref{Eq:ModifiedHKY}
will be referred to as
the HKY+IS-IGC model and will be contrasted to the HKY+PS-IGC model
that combines HKY-IND with the pair-site (PS)
IGC treatment that is described below.

The PS approach jointly considers corresponding sites from
all paralogs in the same genome in a pairwise manner.
When there are two paralogs, the PS approach jointly considers
the states of four nucleotides (two sites from each of the two paralogs).
This transforms a 4-state nucleotide substitution
model into a $4^4 = 256$-state
joint nucleotide substitution model.
Consider the site at position $a$ and the site
at position $b$ ($a < b$) in paralog $i$.
The states of these sites will be denoted $i_a$ and $i_b$.
The corresponding states at positions $a$ and $b$ of
paralog $j$ will be $j_a$ and $j_b$.
We define $Q_{(i_a,i_b,j_a,j_b),(i_a',i_b',j_a',j_b')}$  to be the
instantaneous rate at which $i_a$ changes to $i_a'$
and $i_b$ changes
to $i_b'$ while $j_a$ changes to $j_a'$ and $j_b$ changes to $j_b'$.
An IGC event involving one or both of the sites must yield
$i_a' = j_a'$ or $i_b' = j_b'$ or both.  This homogenization is
reflected in the
rates of the HKY+PS-IGC model by considering how often
IGC events affect both positions $a$ and $b$ in the two paralogs
and how often they affect just one of positions $a$ and $b$.
The HKY+PS-IGC model
adds these IGC contributions to the rates described by
the above HKY-IND model.


The rates at which IGC events simultaneously affect both positions $a$ and $b$
depend on the geometric length distribution of fixed IGC tracts.
Specifically, assume positions $a$ and $b$ are separated by $n$ sites.
The rate of IGC events that affect site $a$
but not site $b$
is $\frac{\eta }{p}\left[ {1 - {{\left( {1 - p} \right)}^n}} \right]$.
Likewise, the rate of IGC events that affect site $b$ but
not site $a$
is $\frac{\eta }{p}\left[ {1 - {{\left( {1 - p} \right)}^n}} \right]$.
The rate of IGC events that simultaneously affect both sites $a$ and $b$ is $\frac{\eta }{p}{\left( {1 - p} \right)^n}$.
This means that the evolutionary dependence between sites $a$ and $b$ due to
IGC is a function of the separation between $a$ and $b$ along
the sequence
and becomes weak when ${\left( {1 - p} \right)^n}$ is near $0$.

Changes that can only be caused by point mutations have the rates
defined only from the rates of the HKY-IND model
(e.g., ${Q_{(A,C,G,T),(C,C,G,T)}} = {Q_{AC}}$).
Changes that can be caused by either one point mutation event or
by
an IGC event that covers only one of the two sites have the rates
defined by the sum of the HKY-IND point mutation rate and
the IGC rate (e.g.,
${Q_{(A,C,G,T),(G,C,G,T)}} = {Q_{AG}} +
\frac{\eta }{p}\left[ {1 - {{\left( {1 - p} \right)}^n}} \right]$).
Changes that
can be caused by either one point mutation event or
by an IGC event that affects either
only one or both of the two sites
have the rates defined by the sum of the HKY-IND point mutation rate
and the two IGC rates
(e.g., ${Q_{(A,T,G,T),(G,T,G,T)}} = {Q_{AG}} + \frac{\eta }{p}$).
Because the instantaneous rate of point mutations originating at two sites that
are both destined for fixation is assumed to be negligible, changes that
simultaneously modify both sites
are exclusively determined by the IGC contribution
(e.g., ${Q_{(A,C,G,T),(G,T,G,T)}} =
\frac{\eta }{p}{\left( {1 - p} \right)^n}$).
The off-diagonal entries of this rate matrix that may be non-zero therefore
have this structure:

\begin{equation}
Q_{(i_a,i_b,j_a,j_b),(i_a',i_b',j_a',j_b')} = \left\{ {\begin{array}{ll}
Q_{i_a,i_a'} \quad & {\mbox{if } i_a \ne i_a',i_b = i_b',j_a = j_a' \ne i_a',j_b = j_b'}\\
Q_{i_a,i_a'} + \frac{\eta }{p}\left[ {1 - (1 - p)^n} \right] \quad &{\mbox{if }i_a \ne i_a',i_b = i_b',j_a = j_a' = i_a',j_b = j_b' \ne i_b'}\\
Q_{i_a,i_a'} + \frac{\eta }{p} \quad &{\mbox{if }i_a \ne i_a',i_b = i_b', j_a = j_a' = i_a', j_b = j_b' = i_b'}\\
{\frac{\eta }{p}(1 - p)^n} \quad & {\mbox{if }i_a \ne i_a',i_b \ne i_b', j_a = j_a' = i_a', j_b = j_b' = i_b'}\\
{...}&{}
\end{array}} \right.
\label{Eq:HKY+PS-IGC}
\end{equation}
Although Equation~\ref{Eq:HKY+PS-IGC} only details rates for changes that alter site $a$ in paralog
$i$,
other rates can be derived similarly.
In summary,
instantaneous rates can be positive only if the corresponding change could
be caused by a single point mutation or a single IGC event or both.

To infer parameter values, we follow \cite{McVean2002} by
forming a composite likelihood
that is the product over all possible pairs of sites of
pairwise marginal likelihood. Let $s_1, s_2, ..., s_N$ represent
the $N$ columns in a multiple sequence alignment.
For the situation where there are exactly two paralogs
(paralog $i$ and $j$),
each alignment
column will be assumed to specify the corresponding states of both
paralog $i$ and paralog $j$ from all species
being considered. Also, a column may have
corresponding nucleotides from outgroup taxa that diverged
prior to the time when paralogs $i$ and $j$
were formed by the duplication.
Parameter values are estimated by
maximizing the composite likelihood.
Specifically, the branch lengths and rates in the model will be denoted
by the vector $\theta$.  The maximum composite likelihood estimate (MCLE)
of $\theta$
is therefore
\begin{equation}
{\hat \theta _{MCLE}} = \argmax_{\theta} \prod\limits_{1 \le a < b \le N} {\Pr ({s_a},{s_b}|\theta )},
\label{Eq:Objective}
\end{equation}
where each pairwise marginal likelihood is calculated with Felsenstein's pruning
algorithm \citep{Felsenstein1981} and where Equation~\ref{Eq:HKY+PS-IGC}
determines rates for portions of the
species phylogeny that have two paralogs present and
Equation~\ref{Eq:ModifiedHKY} specifies
rates when only one paralog is present.

Our inference approach treats
the complex correlation between multiple sites with a pairwise
composite likelihood.
As reviewed in \cite{Varin2011}, maximum composite likelihood estimates of pairwise composite likelihood functions are asymptotically unbiased.
This means that parameter estimates should approach their true values as paralog
length increases.
For simple cases where the full likelihood could be calculated
in their study of homologous recombination,
\cite{McVean2002} showed
that their composite likelihood
estimate was close to the maximum (full) likelihood estimate but
with variance exceeding that of the full likelihood. Our
PS-IGC approach is a pairwise composite
likelihood method that is similar in spirit to the one of \cite{McVean2002}
and should have qualitatively similar behavior.

Since it is
not a valid likelihood function, conventional asymptotic variance
calculations via the Fisher information
matrix are not applicable to the PS-IGC approach.
The uncertainty of the PS-IGC parameter estimates
is approximated in this study through the parametric bootstrap (e.g., see
\citealt{Goldman1993}).
A less computationally demanding
alternative might be to approximate
the uncertainty via the inverse of the Godambe information matrix
(e.g., see \citealt{Kent1982}; \citealt{Varin2011}).

\section{{Results} \label{sec:Results} }

We analyzed both actual and simulated data sets.  Information
regarding
the software implementation and the numerical optimization scheme are provided
in the Materials and Methods.

\subsection{Analyses of Simulated Data}

\begin{figure*}[t]
\begin{center}
\includegraphics[scale=0.8]{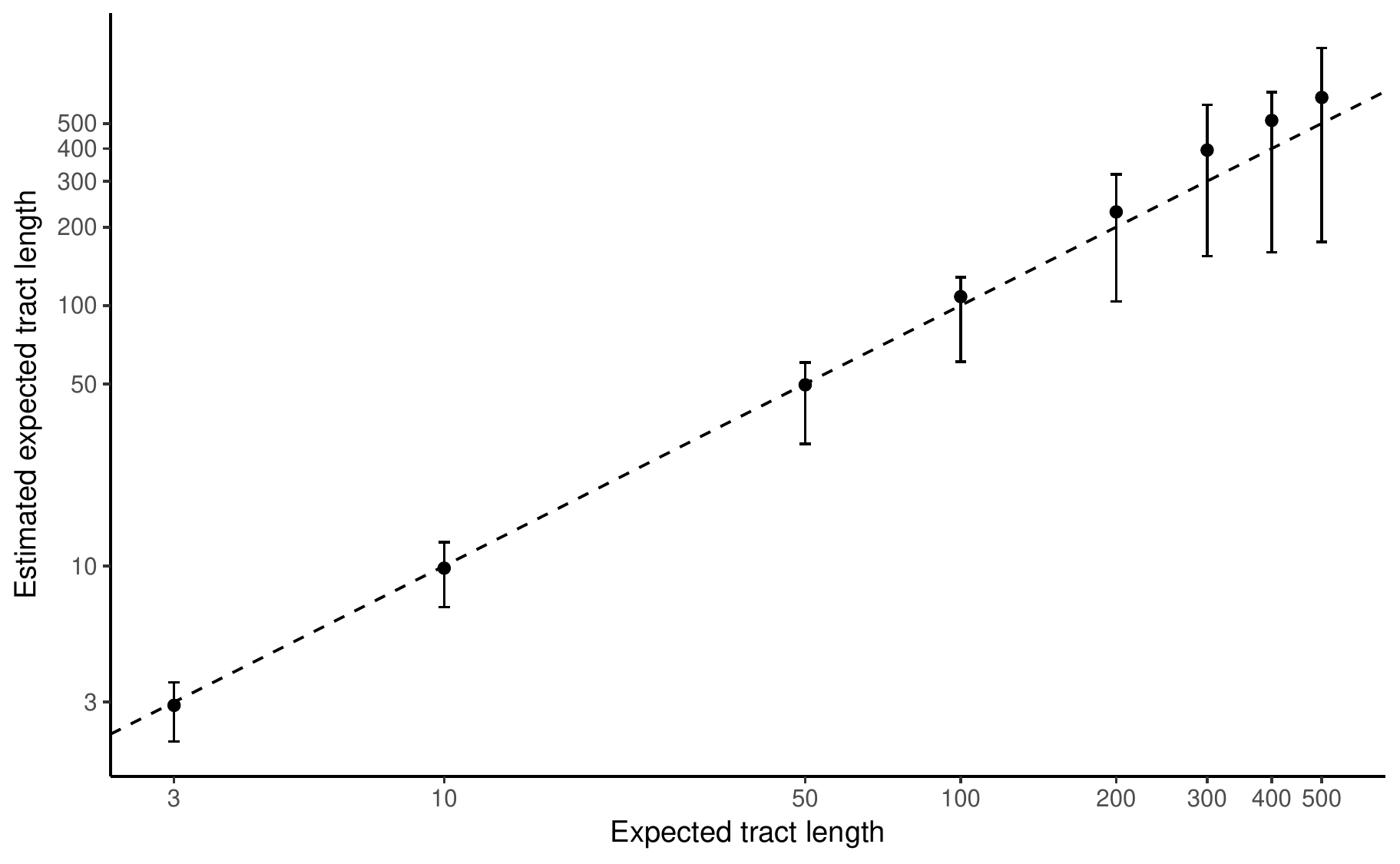}
\end{center}
\caption{Expected IGC tract length
versus estimates with maximum composite likelihood from simulated data.
All simulation scenarios used a value of $\tau = \eta / p = 5.16$.
One hundred data sets were simulated for each
true value of $1/p$.
As described in the text, maximum composite
likelihood estimates of $1/p$ 
that exceeded the true value by a factor of $10$ or more
were excluded for the sample mean calculation but were included for the
reported interquartile range values.
The x-axis shows the true value of the expected IGC tract length (i.e.,
$1/p$) and the y-axis shows sample means among estimates with vertical line
segments depicting interquartile ranges of the estimates.
The dashed line shows $Y=X$.
}
\label{fig:PSIGCSimulationStudy}
\end{figure*}

Simulations were performed
to characterize our pair-site composite likelihood IGC procedure
as detailed in Materials and Methods.
Figure \ref{fig:PSIGCSimulationStudy}
summarizes estimates of the average fixed tract length (i.e., $1/p$)
from these simulations.
Sometimes, simulations yielded inferences for $1/p$
(i.e., estimated average tract lengths) that were more than 10-fold higher
than the true value. These extreme values were not included in sample mean
calculations but they were included in the reported interquartile ranges.
No estimated values were discarded for expected
tract lengths of 3, 10, 50 and
100 whereas 3, 7, 6 and 7 estimated values were respectively discarded for
the expected tract lengths of 200, 300, 400 and 500.

With the important caveat that not all parameters were estimated from
simulated data (see Materials and Methods),
the simulations indicate that average estimated mean tract lengths
are relatively close to the true values
but
the variability of expected tract length estimates increases as expected
tract length increases.  Presumably, this correlation
is partially attributable
to the fact that all simulation scenarios share the same value for the
product of the IGC tract initiation rate and the expected tract length.
Therefore,
the actual numbers of IGC
events will vary more among simulated data sets when tracts are long
but the expected number of tracts per simulation is small.

\subsection{Analyses of Actual Data}

We applied the PS approach to analyze three groups of datasets.
The first group consists of the $14$ data sets that \cite{Ji2016}
analyzed with their codon-based IS approach for studying IGC.  The second
group is actually a single primate data set that \cite{Zhang1998}
considered in their pioneering work on the origin of gene function.
The third group of data sets are segmentally-duplicated intron regions from
primates. \cite{Harpak2017} studied the effects of IGC on these intron
regions with their binary-state treatment and we chose these data to
examine how
IGC inferences are affected by instead using maximum composite likelihood
with a nucleotide-based model.


\subsubsection{Yeast Paralogs} \label{sec:YeastResults}
\begin{figure*}[t]
\begin{center}
\includegraphics[scale=0.3]{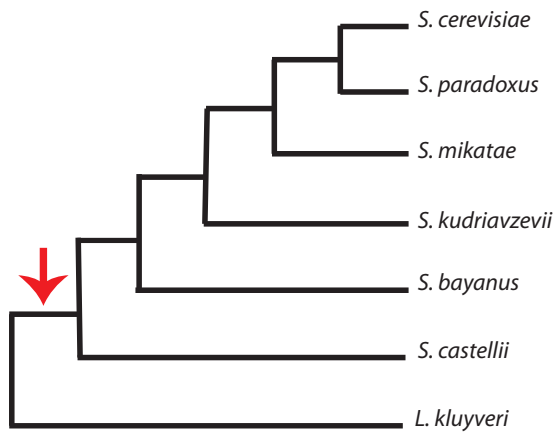}
\end{center}
\caption{The tree topology used for evolutionary analyses of the
yeast datasets. The arrow indicates the
branch on which the genome-wide duplication occurred. }
\label{fig:YeastSpeciesTree}
\end{figure*}

Yeast experienced an ancient genome-wide duplication
(\citealt{Wolfe1997}; \citealt{Philippsen1997}; \citealt{Kellis2004}; \citealt{Dietrich2004}; \citealt{Dujon2004}).
\cite{Ji2016} analyzed
14 data sets of yeast protein-coding genes
to characterize IGC that occurred subsequent to the genome-wide duplication.
These data sets were the only ones that remained after applying
stringent filters that were designed
to reduce concerns about sequence alignment and paralogy status
(see \citealt{Ji2016}).  While the filters did not require
it, all 14 data sets happen to encode ribosomal proteins.
In every data set, six yeast species are each represented by two paralogs
that stem from the ancient genome-wide duplication.
Each data set also includes a sequence from a species ({\it L. kluyveri})
that diverged from the other six prior to the genome-wide duplication.
The species represented in these data sets are related by
the well-established phylogenetic tree topology of
Figure~\ref{fig:YeastSpeciesTree}.

Table~\ref{tab:IS-IGCYeastResults}  summarizes some
of the results obtained by analyzing the 14 yeast
data sets with the HKY+IS-IGC model.  As was the case when these data
sets were analyzed with the codon-based model of
\cite{Ji2016}, Table~\ref{tab:IS-IGCYeastResults}
shows that a substantial proportion of sequence change
is attributed to IGC by both the HKY+IS-IGC and HKY+PS-IGC models.
Table~\ref{tab:IGCestimates}
contrasts other inferences from the HKY+IS-IGC and HKY+PS-IGC models
for these 14 data sets.  It shows both models yield very similar estimates
of $\tau$.
Also,
Table~\ref{tab:IGCestimates} reveals
that the expected fixed
IGC tract lengths tend to be quite short according to the
HKY+PS-IGC estimates.  In fact, only 1 of the 14 data sets yields an expected
tract length that exceeds 100 nucleotides and 8 of the 14 data sets
yield expected tract lengths that are less than 20 nucleotides.

\subsubsection{Primate EDN and ECP}

\begin{figure*}[t]
\begin{center}
\includegraphics[scale=0.2]{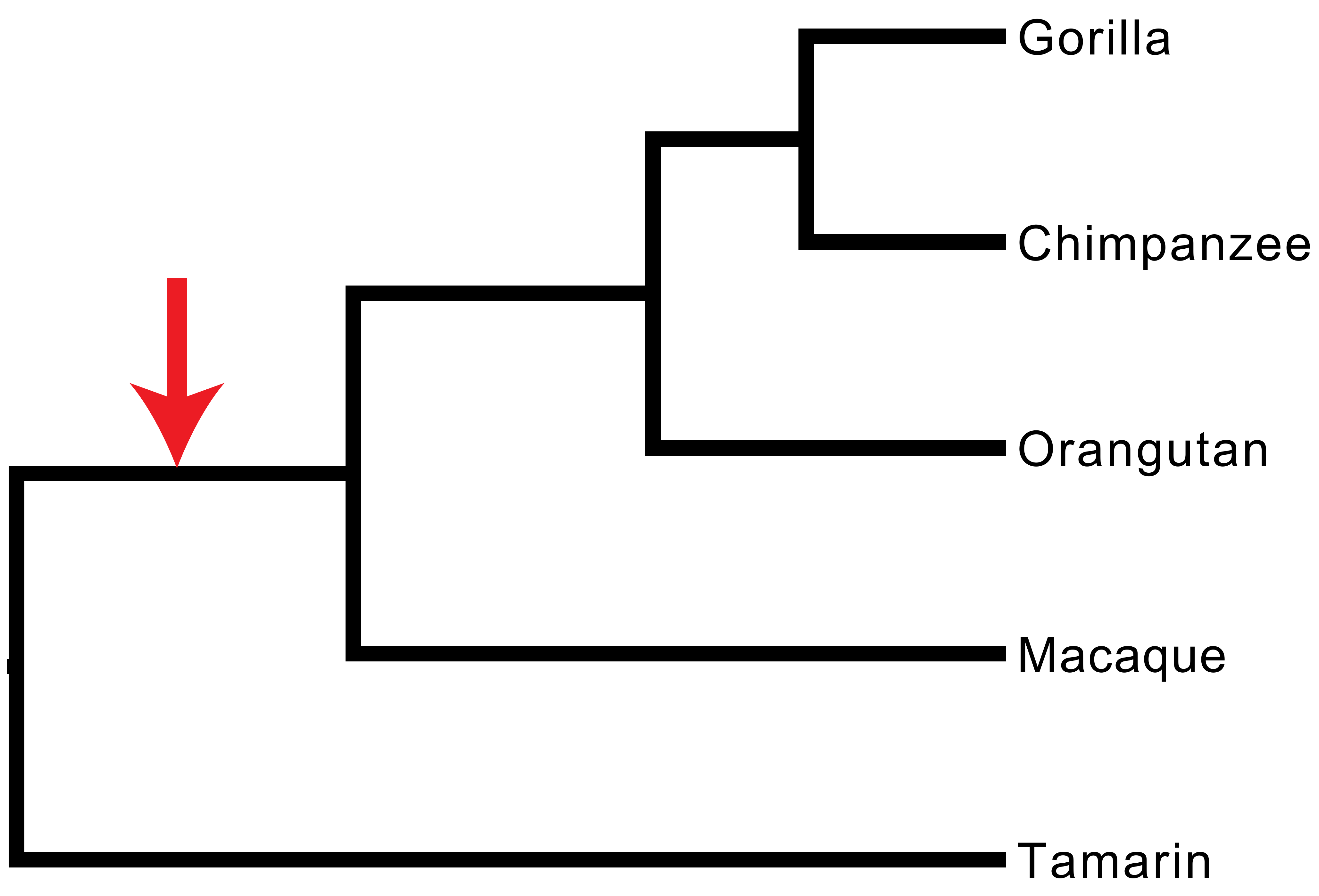}
\end{center}
\caption{The tree topology used for evolutionary analysis of primate
EDN
and ECP genes. The arrow indicates the duplication event.
}
\label{fig:PrimateSpeciesTree}
\end{figure*}

\cite{Zhang1998} studied primate paralogs that encode
eosinophil cationic protein (ECP)
and eosinophil-derived neurotoxin (EDN).
As described in the Materials and Methods, we used the sequence data and
tree topology (see Figure~\ref{fig:PrimateSpeciesTree}) of \cite{Zhang1998}.

We analyzed these data with both the IS and PS frameworks.  Whereas our PS implementation has the
advantage of accounting for IGC tracts, our IS implementation is
computationally
feasible with codon-based substitution models.
Using an adaptation of the Muse-Gaut codon
model \citep{Muse1994}
that we will denote the MG94+IS-IGC model
(see \citealt{Ji2016})
and considering the changes that occurred subsequent to the EDN/ECP duplication,
we estimate that approximately 10.3\% of the codon substitutions originated
with an IGC
event rather than a point mutation.

\begin{figure*}[t]
\begin{center}
\includegraphics[scale=0.6]{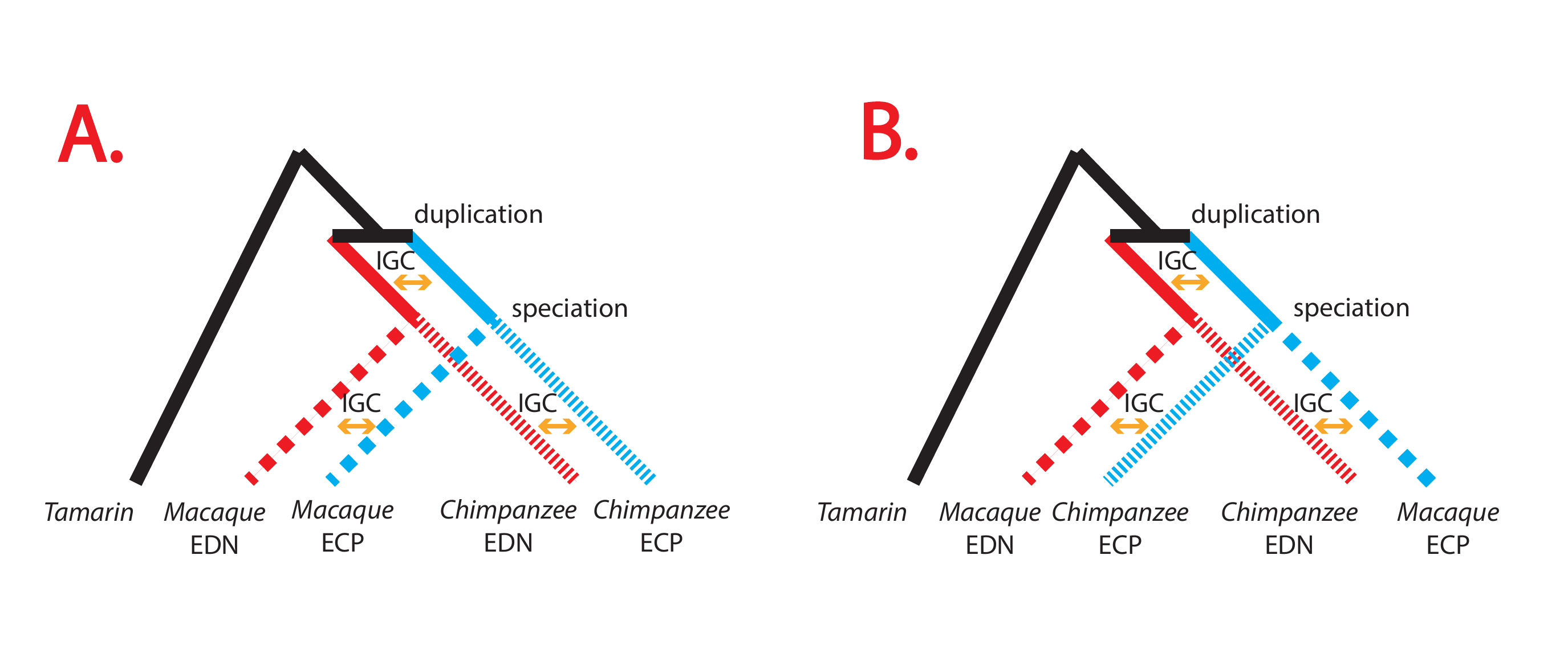}
\end{center}
\caption{
A paralog-swapping experiment addressing whether improvement to model fit can be attributed to IGC
or to artifacts. Both Scenarios $A$ and $B$ specify the correct phylogeny between Tamarin
and the paralogs of Macaque and Chimpanzee. Scenario $A$ shows the biologically correct situation that has
 the IGC between paralogs in the same genome.
 In Scenario $B$, IGC homogenization events involve one paralog in Macaque and one from Chimpanzee.
 Because Scenario $A$ corresponds to how observed data are generated, Scenario $A$ should fit better than
 Scenario $B$ if IGC is actually being detected. Note that this paralog-swapping experiment would not be
 possible if only $1$ postduplication species was used and would not be effective with more than
 two postduplication species.}
\label{fig:EDN_ECP_SwapTest}
\end{figure*}

In \cite{Ji2016}, we introduced a ``paralog-swapping''
procedure that is intended to
investigate whether inferred IGC levels are not actually due to IGC events but are instead
attributable to estimation artifacts that arise
because of imperfect evolutionary models.
The paralog-swapping procedure
uses the two paralogs from each of two taxa that are descended from
a post-duplication speciation event.
It compares the biologically plausible scenario
where IGC involves paralogs in the same genome to a biologically implausible
scenario that has IGC between paralogs in different genomes
(see Figure~\ref{fig:EDN_ECP_SwapTest}).
The idea underlying the paralog-swapping procedure is that the inferred
IGC levels will be similar between the two scenarios if artifacts due to unrealistic
evolutionary models are generating the putative IGC signal.
When we apply the paralog-swapping procedure in conjunction with the HKY+IS-IGC model,
we infer much more IGC with the biologically plausible scenario.
With the biologically plausible scenario of Figure~\ref{fig:EDN_ECP_SwapTest}A,
the IGC parameter $\tau$ is estimated to be about $0.28$.  With the biologically
implausible scenario of Figure~\ref{fig:EDN_ECP_SwapTest}B,
the IGC parameter $\tau$ is estimated to be about
$0.00004$ (i.e., very close to $0$).

Table~\ref{tab:IS-IGCYeastResults} includes
results obtained by analyzing the EDN/ECP
data set with the HKY+IS-IGC model.  While the proportion of
sequence change attributable to IGC is smaller for the EDN/ECP
data set than any of the 14 yeast data sets, Table~\ref{tab:IS-IGCYeastResults}
shows that this inferred proportion is substantial whether the EDN/ECP
data are analyzed with the HKY+IS-IGC model or the codon-based IGC treatment
of \cite{Ji2016}.
As was the case for most of the yeast data sets, Table~\ref{tab:IGCestimates}
shows that the fixed IGC tract lengths that are inferred from EDN/ECP
are very short.

%

%

\subsubsection{Primate Introns}

\cite{Harpak2017} examined the effects of IGC on the evolution of
segmentally-duplicated primate introns.  Here,
we analyze a subset of the same data
set in order to characterize IGC with our more parameter-rich models.
As described in the Materials and Methods,
we further filtered the introns of \cite{Harpak2017}
to yield 20 data subsets where
all ingroup species have two paralogs and the outgroup species
has one paralog.
This was done to lessen uncertainty regarding paralogy versus orthology.
This filter also serves to make a single branch on the rooted primate tree
be where the
duplication event likely occurred for all 20 data subsets.

\begin{figure*}[t]
\begin{center}
\includegraphics[scale=0.2]{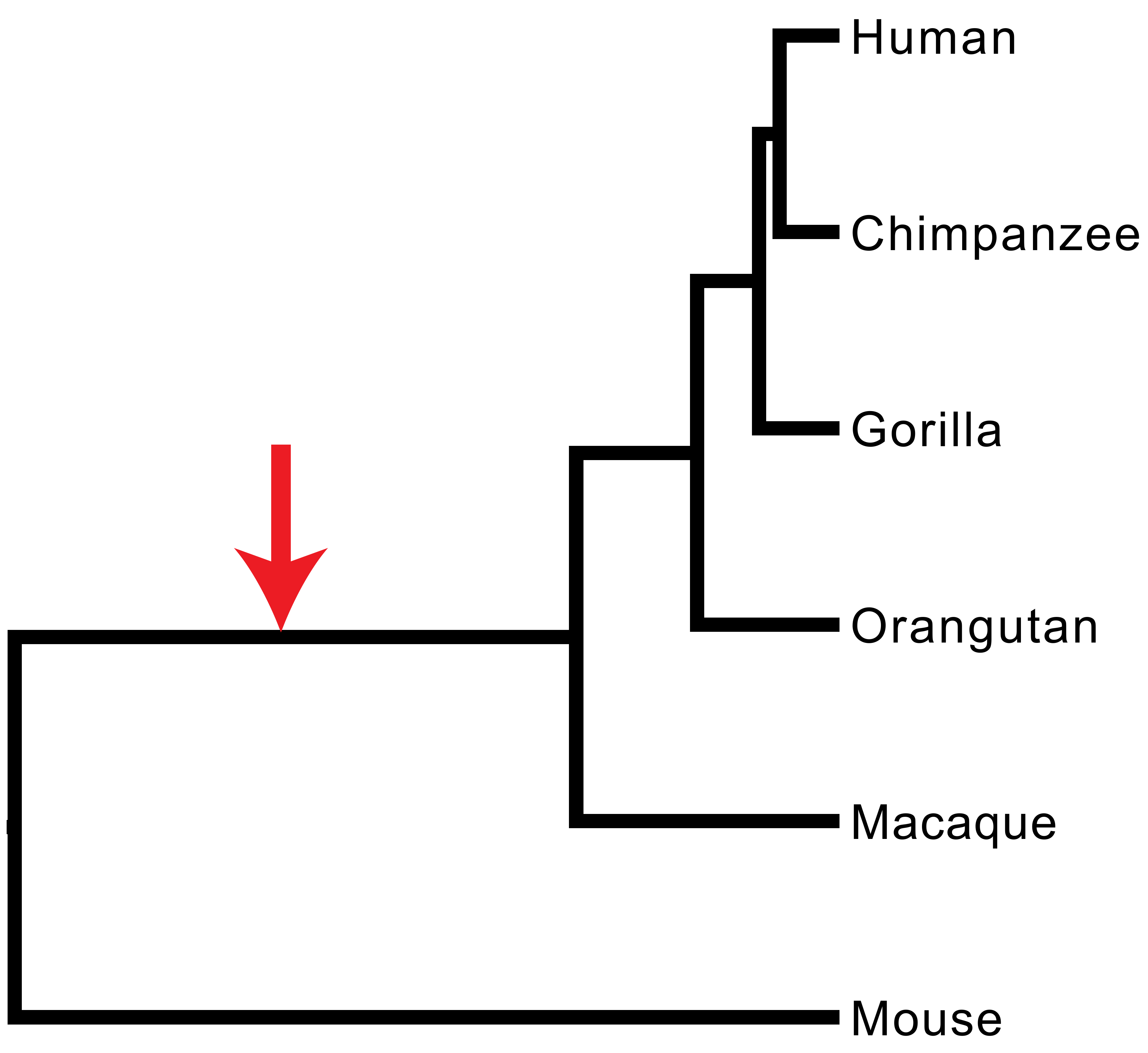}
\end{center}
\caption{The tree topology used for evolutionary analysis of the primate intronic sequences. The arrow indicates the segmental duplication event.
}
\label{fig:HarpakTree}
\end{figure*}


Table~\ref{tab:IS-IGCYeastResults} includes
results obtained by analyzing the primate introns
under the HKY+IS-IGC model that has no rate heterogeneity (i.e. $r_2 = r_3 = 1$).  It shows that the proportion of
sequence change attributable to IGC is smaller for the primate
intron data than for EDN/ECP
and than for each of the 14 yeast data sets, but this proportion
is still relatively large.
In contrast to the results from the yeast and EDN/ECP protein-coding
data, Table~\ref{tab:IGCestimates}
shows that the fixed IGC tract lengths that are estimated
from the intron data have expected values
that exceed several hundred nucleotides.

\section{{Discussion}\label{sec:Discussion} }

\subsection{Fixed IGC tracts inferred to be short}

A distinction can be made between the lengths of IGC mutations and the lengths
of IGC tracts that experience fixation.
Measurements of the lengths of IGC mutations vary widely among
studies and model systems, with some events affecting only about
$10$ nucleotides and with more typical lengths involving hundreds
of consecutive positions (e.g., see
\citealt{Chen2007,Mansai2011}).
Our estimates of the mean lengths of fixed IGC tracts also exhibit substantial
variation (see Table~\ref{tab:IGCestimates}).
Whereas our estimates of the mean fixed IGC tract lengths from intron
data tend to be not
dramatically different than tract length estimates for IGC mutations,
our estimates from exon data of the lengths of fixed IGC tracts
(see Table~\ref{tab:IGCestimates})
are substantially smaller
than are usually obtained from studies
of IGC mutations.
This disparity may be partially attributable
to
homologous recombination events that occur
subsequent to IGC mutation
and that could thereby make the
lengths of fixed IGC tracts shorter than
the length of the original IGC mutations.

Natural selection may enhance the disparity between lengths of IGC mutation
tracts and lengths of fixed tracts.  IGC mutations can
simultaneously introduce multiple fitness-affecting sequence changes into
a paralog.  Some of the changes may be advantageous whereas others may be
deleterious.  Subsequent homologous recombination may separate the
advantageous and disadvantageous changes and thereby favor the fixation of
IGC tracts that are shorter than the lengths of the original IGC mutations.

Biologically implausible evolutionary models represent a more mundane
explanation for the short lengths that we estimate for fixed exonic IGC
tracts.
One shortcoming of the analyses
presented here
is that our treatment of changes due to IGC is more unrealistic
than the treatment of substitutions that originated with point mutation.
For nucleotide substitutions that originated by point mutation,
the pair-site approach reflects the difference of fixation rates
among the three codon positions by setting the relative rate at the
first codon position to $1$ and adding a separate relative rate parameter
for the second position
(e.g., $r_2$) and another for the third position (e.g., $r_3$).
However, the pair-site approach does not have differential fixation rates
according to which codon positions are affected by the IGC event.
This IGC treatment is convenient for setting up
the statistical model but may be biologically unrealistic.
Furthermore, we assume that IGC mutations occur and fix at a rate
that is independent of the differences between the IGC donor and
recipient tracts. This assumption violates evidence
that IGC mutations become less likely as paralogs diverge
(see \citealt{Chen2007}).



Whereas the exons analyzed in this study are from genes that
contain either no intron or a short one,
further investigations with more genes and more sophisticated IGC models
may shed light on whether fixed IGC stretches really are shorter in exons
than introns.
These future investigations might benefit from studying
protein-coding genes with both
long coding regions and long introns.

\subsection{IGC inference procedures}

\cite{Harpak2017} recently developed two inferential procedures so that
they could study IGC in segmentally-duplicated
regions of humans and other primates.
One of these procedures employed a hidden Markov model (HMM) that relied on the
assumption that each sequence site has experienced either $0$ or $1$ IGC
events subsequent to the duplication that created the paralogs being studied.
Based on analyses of simulated data,
we found a closely related but distinct HMM procedure
to yield misleading estimates of expected IGC
tract lengths, possibly because the IGC levels that were examined
caused serious violations of the HMM assumptions \citep{Ji2017}.
HMM approaches have proven useful for diverse DNA and protein
sequence analysis
tasks, but their value for studying IGC is likely to be particularly
sensitive to how closely aligned are the HMM assumptions to the data
being analyzed.

The other inferential procedure explored by \cite{Harpak2017} was a pair-site
IGC treatment that resembles the one
introduced here (see also \citealt{Ji2017}).
An attractive feature of the pair-site approaches is that they employ maximum
composite likelihood for inference rather than relying on HMM assumptions that
are most justified when IGC is rare.
The pair-site approach of \cite{Harpak2017}
considers two sites and two paralogs jointly but treats all sequence
sites as being binary in nature. As a result, their joint state space
has a size of $2^4 = 16$ where each site has two possible states
representing two allele types with one state being the nucleotide
type observed in mouse and with the other binary state collectively
representing the other $3$ nucleotide types.
Furthermore,  \citeauthor{Harpak2017}
assume constant generation time and known divergence
times on the lineage separating the primates from their most recent
common ancestor with mouse. The \citeauthor{Harpak2017} pair-site approach seems
most appropriate for evolutionary scales where sequence changes are rare
and where species are relatively closely related so that parameters
such as generation time have little variation across the tree.
It may be less satisfactory for larger timescales, such as the one
that encompasses the period from the common ancestor of rodents
and primates to the current day.
In contrast, our parameterizations have been aimed at
timescales of this magnitude.

\citeauthor{Harpak2017} concluded from their intron analyses
that the IGC rate per sequence position
is
an order of magnitude faster than the point mutation rate.
Because the rates of the IS-IGC and
PS-IGC approaches are normalized so that
the expected rate per paralog per site is $1$ for substitutions
that originated with a point mutation, our estimated $\tau$ values of
approximately $0.5$ from both the IS-IGC and the PS-IGC approaches
suggest that IGC happens at roughly the same rate per site of each
paralog as point mutation.
In addition, we estimate the percentage of nucleotide substitutions that
originate with an IGC event rather than a point mutation to be $9.3\%$ for
the filtered
intron datasets.
The estimated average tract length from our PS-IGC approach has the same order
of magnitude as estimated by \citeauthor{Harpak2017}.
Because of the substantial differences between our analyses
and those of \citeauthor{Harpak2017}, it is unclear how to isolate the cause
of the different results.

\subsection{Abundant substitution due to IGC}

Following \cite{Ji2016}, we used the MG94+IS-IGC model
to estimate the proportion of fixed codon changes that
were attributable to IGC
rather than point mutation.
We also employed the HKY+IS-IGC model to infer the proportion
of nucleotide changes that were attributable to IGC rather than point mutation.
For both the Yeast data sets and the primate EDN/ECP data, the
estimated proportions are somewhat higher
for the HKY+IS-IGC model than for the MG94+IS-IGC model (see Table
\ref{tab:IS-IGCYeastResults}).
While we expect the proportion for the HKY+IS-IGC model to be somewhat
higher than for the MG94+IS-IGC model because individual events that change multiple
codon positions are each only counted once in the MG94+IS-IGC calculations, we expect that
most of the disparity in proportions is due to the differences between models.

While there is some disparity in IGC estimates among procedures,
the more important message from Table
\ref{tab:IS-IGCYeastResults} is there is a substantial amount of evolutionary
change that is due to IGC.  This has implications for the evolutionary
consequences of gene duplications.
Gene duplication is considered an important source of novel gene function.
After their formation, duplicated genes may experience
neo-functionalization, sub-functionalization or become pseudogenized (e.g., see \citealt{Lynch2000, Walsh2003}).
In the absence of IGC, these events are determined by mutations accumulating independently at each paralog.
\cite{Teshima2008} showed that IGC can slow down the fixation of the neofunctionalized paralog by
overwriting the neofunctionalized paralog sequence by that of another paralog.  In this way, IGC can oppose natural selection.
Furthermore, consideration of IGC could potentially improve the dating of
subfunctionalization or neofunctionalization events.

Our conclusion is that fixation of IGC mutations is an important source of
evolutionary change in multigene families.  While the relative rate of IGC
experienced per site (i.e., $\tau$)
can be relatively well estimated by our IS and PS
techniques,
our PS approach yielded estimates of mean tract length
from protein-coding data that were unexpectedly short and
that were
associated with high degrees of uncertainty.  This motivates development
of additional inference procedures that might be better than
the PS procedure at extracting IGC tract
information.
In addition, there is ample room for improvement of IGC inference procedures.
One important future direction will be to have the inference
procedures assess how IGC rates decrease with paralog divergence.
In addition, many multigene families consist of
more than two paralogs and methods are needed for characterizing IGC in these
cases.  Because our analyses suggest that IGC is responsible for a substantial
proportion of molecular evolution in multigene families, we believe that
improved IGC inference should be a high priority.

\section{{Materials and Methods} \label{sec:MaterialsMethods}}

\subsection{Rate Normalization}

For all analyses, substitution rates were normalized so that
the expected rate per paralog per nucleotide is $1$ for substitutions
that originated with a point mutation and so that
the reported values of $\tau$ can then be compared to
this normalized value.
The normalization leads to one unit of branch length representing
one expected substitution
arising from point mutation per paralog per site.
Because the model has rate heterogeneity among the three codon positions,
the normalization to an average rate of $1$ yields
respective
expected rates of $3/(1+r_2+r_3)$, $3r_2/(1+r_2+r_3)$, and $3r_3/(1+r_2+r_3)$
at the first, second, and third codon positions.

\subsection{Simulations} \label{ssec:Simulations}

The Materials and Methods of \cite{Ji2016}
describe how and why data sets were simulated using parameter values
estimated from the
YDR418W\_YEL054C data set.
With the same simulation procedure and for the same reasons, we again
based our simulations on the YDR418W\_YEL054C data set.
However, this time we used the inferred values of the HKY+IS-IGC parameters
from the YDR418W\_YEL054C data to simulate data sets.

The estimated value of $\tau = \eta / p$ was $5.16$ for the
YDR418W\_YEL054C
data set.  We wanted to explore how the composite likelihood
estimates of expected tract length $1/p$ varied for different true values of
$1/p$.
Because $\eta = \tau / (1/p)$, we used $\tau = 5.16$ and each
value of $1/p$ that was explored in the simulations
to set the corresponding value of $\eta$.
All other parameters were set at the values inferred by maximum composite
likelihood from the YDR418W\_YEL054C data set.

To match the YDR418W\_YEL054C data,
IGC and point mutation events were simulated according to the yeast
species tree
for sequences of length 492 nucleotides.
While the simulations in \cite{Ji2016} operated at the codon level,
these simulations used nucleotides as the units because
the modified HKY model is a nucleotide substitution model
that has independent evolution at the three codon positions
but has different rates for each codon position.
The nucleotides in
three alignment columns (i.e., columns 238, 239, 240) were removed from each
simulated data set because
the actual YDR418W\_YER054C data has a gap in these columns.
For each simulation condition, 100 data sets were generated.

For each simulation scenario,
one hundred data sets were simulated and analyzed with the HKY+PS-IGC model.
Rather than finding the combination
of parameter values that jointly maximize the composite likelihood,
computational concerns and
numerical optimization difficulties resulted in our instead optimizing
the composite likelihood only over the tract length parameter
$p$ with all other free parameters constrained at the values that were used
to simulate the data sets.
We note that the parameter $\tau = \eta / p$.
By constraining $\tau$ at its true value and then inferring $p$, a value
for the IGC tract initiation parameter $\eta$ is simultaneously inferred
(i.e., $\eta = \frac{\tau }{(1/p)}$).
While the computational shortcut of only inferring $p$ from simulated data is
not ideal,
the simulation results of \cite{Ji2016} indicated that the value of $\tau$
can be estimated relatively well.

\subsection{Parametric Bootstrap Analyses}

There were 100 replicates per parametric bootstrap analysis.  The maximum composite
likelihood procedure was employed to obtain estimates of model parameters from
actual data and those estimated values were used to simulate data sets of the
same size as the actual data.  Each of the 100 simulated data sets was analyzed
by the composite likelihood procedure and variability of the resulting estimates
was summarized.  Paralleling the analyses of simulated data sets described in
Section \ref{ssec:Simulations}, analyses of the 100
bootstrap replicates with the HKY+PS-IGC
model were made computationally feasible
by estimating the IGC tract length parameter $p$ but constraining all other free
parameters to their true values.  This imposition of constraints will presumably
cause the uncertainty in the tract length parameter $p$ to be underestimated, but we hope that
this underestimation is small given that $\tau = \eta /p$ and that
\cite{Ji2016} found that the $\tau$ parameter
of the IS model could usually be well estimated.

\subsection{Yeast Data}

In the Materials and Methods of \cite{Ji2016},
we describe how the $14$ data sets of yeast protein-coding genes
were selected and prepared.
Although the 14 yeast data sets exclude introns,
our pair-site analyses of these data attempted to accommodate
the fact that an IGC tract may cross exon-intron boundaries.
For each pair of {\it S. cerevisiae} paralogs that is listed in
Table~\ref{tab:IS-IGCYeastResults} and Table~\ref{tab:IGCestimates},
we identified the exon-intron boundaries
and intron lengths of the first of the listed {\it S. cerevisiae} paralogs.
For each adjacent pair of columns in our alignment of yeast exons, we could
therefore determine the separation along the gene sequence of the
{\it S. cerevisiae} paralog that is listed first for each pair in
Table~\ref{tab:IS-IGCYeastResults} and Table~\ref{tab:IGCestimates}.
We then assumed that the gene sequence separation between adjacent exon
columns for this {\it S. cerevisiae} paralog was identical to the gene sequence
separation for all other paralogs in the exon alignment.  Therefore, we did not
attempt to account for insertions and deletions or for
the possibility that exon-intron boundaries
may vary along the evolutionary tree.

\subsection{Primate EDN and ECP Data}

We used the EDN and ECP sequence
data from the \cite{Zhang1998} study and the tree topology
that was first introduced by \cite{Rosenberg1995}.
Because of the relatively short time separating the common
ancestor of humans and chimpanzees from the common ancestor of those
two species with gorillas, we excluded human sequence data from our analyses
to lessen the variation of gene tree
and species tree topologies.
 The remaining species are gorilla ({\it Gorilla gorilla}),
chimpanzee ({\it Pan troglodytes}), orangutan ({\it Pongo pygmaeus}), macaque ({\it Macaca fascicularis}), and tamarin ({\it Saguinus oedipus}).
We obtained the protein-coding DNA sequences via their
GenBank accession numbers
and aligned them at the amino acid level with version 7.305b
of the MAFFT software \citep{Katoh2013}.
The protein sequence alignment was then converted to
the corresponding codon-level alignment.
Three codon columns that contain gaps were removed from the alignment.
The remaining $157$ codon columns were used in the analyses.

\subsection{Primate Intronic Data}

We received $550$ sequence alignment files of the intronic data
analyzed from \cite{Harpak2017}.
Each alignment file corresponds to one intron of a total of $75$ protein-coding gene pairs.
We further filtered the alignment files to retain
only those with both paralogs in the five ingroup species:
human ({\it Homo sapiens}), chimpanzee ({\it Pan troglodytes}), gorilla ({\it Gorilla gorilla}), orangutan ({\it Pongo abelii})
and macaque ({\it Macaca mulatta}); and one paralog in the outgroup
species: mouse ({\it Mus musculus}).
This filter is intended
to minimize uncertainties of gene duplication and loss histories
so that the analyzed data is less likely to
violate the gene tree topology that was assumed in our analyses.
Twenty sets of paralogous introns
and $15370$ sites remained after this
filtering of alignment files.
Each of these twenty sets of paralogous introns
corresponds to one of five pairs of
paralogs of protein-coding genes.
The ensembl gene ids for the paralog pairs from human are: ENSG00000109272
and ENSG00000163737,
ENSG00000136943 and ENSG00000135047, ENSG00000158485 and ENSG00000158477,
ENSG00000163564 and ENSG00000163563, and\\
ENSG00000187626 and ENSG00000189298.

Because these sequence data represent introns rather than exons, we
analyzed the 20 datasets with parameter values under the HKY+PS-IGC model that has no rate heterogeneity (i.e., $r_2 = r_3 = 1$).
In all analyses of the segmentally-duplicated intron data, we assumed that IGC
tracts affecting one pair of intron paralogs in our data set
would not also affect other pairs in our filtered data
set.
While this treatment of exon-intron boundaries is less sophisticated than that
applied to the yeast and EDN/ECP data (see below), it
is computationally less demanding
than the treatment applied to those data.

We investigated two different sorts of analyses of the intron data.  One
sort consisted of grouping together all introns associated with the same
gene.  This results in the 20 intron subsets being divided into 5 groups.
We separately analyzed each of the 5 groups.  We constrained each intron
assigned to a group to share parameter values with the other introns in the
group.  This treatment caused numerical optimization difficulties
for most of the groups.  Specifically, substantially different
maximum composite likelihood estimates were obtained when numerical
optimization was initiated from different
sets of parameter values.
We believe this difficulty arose because most of the 5 groups of introns
contained little information about IGC tract lengths.

Due to this difficulty, we decided to share information across the
5 groups of introns by jointly analyzing them
with the assumption that all introns shared the same parameter values.
This corresponds to an assumption that the duplications that generated
the 5 groups of introns all occurred at the same time.  While this assumption
neglects some variability in branch lengths among gene trees, it results
in sharing information about IGC tract length distributions across the
groups.
Because
this shared treatment resulted in less numerical optimization difficulty,
all parameter estimates reported here for the intron data were obtained via
the shared treatment.

\subsection{Exon-Intron Boundaries}

For yeast and primate EDN/ECP data,
we only analyzed the coding sequences of protein-coding genes but we
incorporated the possibility that individual IGC events could span entire
introns and thereby affect consecutive exons.
Consider two neighboring sites in the coding sequence that
are at positions $k$ and $k + 1$.
Because of the possibility of introns, these two sites may not be neighbors
in the gene sequence. We will assume these sites are separated
in the gene sequence by $m$
nucleotides where $m \ge 1$ with $m=1$ being the case where
the two sites are in
the same exon.
When the two sites
are in different exons, our analyses have
the distance $m$ between them be the length of the
intron sequence plus one.

\subsection{Numerical Optimization}

Model parameters were estimated by numerically optimizing
the logarithms of the composite likelihoods.  Inferred parameters
include branch lengths, parameters related to point mutation,
and IGC-related parameters.
The ``L-BFGS-B'' method from the scipy package
\citep{Oliphant2007,Walt2011} was employed for numerical
optimization.

%

\clearpage

\section{Acknowledgments}
We thank Alexander Griffing, Hirohisa Kishino, Eric Stone, Ed Susko,
and Marc Suchard for diverse assistance.
We also thank Arbel Harpak for sharing his primate intron data.
This work was supported by the National Institute of General Medical Sciences
at the National Institutes of Health (GM118508)
and the National Science Foundation (DEB 1754142).
Data sets are available at
\url{https://github.com/xji3}.    Software for inferring IGC is available at
\url{https://github.com/xji3/IGCexpansion} and
\url{http://jsonctmctree.readthedocs.org/en/latest/}.

\clearpage

\newpage
\begin{table}[h]
\caption{HKY+IS-IGC results.}
\centering
\begin{tabular}{lccccccc}
\toprule
& & & & & & {\bf HKY} & {\bf MG94}\\
{\bf Paralog Pair} & {\bf LnL} & {\bf Diff } & {\bf $\tau$ } & {\bf $r_2$} & {\bf $r_3$} & {\bf Prop} & {\bf Prop}\\
\hline
YLR406C,YDL075W & -1189.81 & 26.55 & 5.10 & 0.43 & 8.08 & 0.24 & 0.20\\
YER131W,YGL189C & -1216.91 & 31.58 & 5.27 & 0.97 & 19.60 & 0.27 & 0.20 \\
YML026C,YDR450W & -1368.47 & 94.72 & 12.85 & 0.27 & 9.67 & 0.40 & 0.34 \\
YNL301C,YOL120C & -2126.64 & 129.75 & 7.94 & 0.57 & 7.16 & 0.32 & 0.26 \\
YNL069C,YIL133C & -2332.61 & 74.95 & 3.63 & 0.43 & 4.78 & 0.26 & 0.22 \\
YMR143W,YDL083C & -1217.38 & 52.80 & 9.19 & 0.13 & 17.72 & 0.31 & 0.29 \\
YJL177W,YKL180W & -1840.38 & 63.64 & 6.45 & 0.50 & 8.97 & 0.28 & 0.21 \\
YBR191W,YPL079W & -1468.95 & 91.84 & 13.66 & 0.15 & 6.39 & 0.40 & 0.32 \\
YER074W,YIL069C & -1233.00 & 131.50 & 20.90 & 0.28 & 6.94 & 0.39 & 0.37 \\
YDR418W,YEL054C & -1735.40 & 65.09 & 5.16 & 0.54 & 11.58 & 0.27 & 0.21 \\
YBL087C,YER117W & -1372.91 & 79.96 & 11.05 & 0.34 & 10.59 & 0.38 & 0.29\\
YLR333C,YGR027C & -1246.67 & 108.84 & 9.88 & 0.60 & 8.05 & 0.39 & 0.29 \\
YMR142C,YDL082W & -2033.88 & 179.32 & 14.37 & 1.12 & 8.42 & 0.42 & 0.38 \\
YER102W,YBL072C & -2037.26 & 205.73 & 14.77 & 0.82 & 6.19 & 0.43 & 0.36 \\
EDN/ECP        & -1713.06 &  12.61 & 1.79  & 1.52 & 1.56 & 0.16 & 0.10\\
Introns & -62878.87 & 93.79 & 0.44  & N.A. & N.A. & 0.09 & N.A. \\
\bottomrule
\label{tab:IS-IGCYeastResults}
\end{tabular}
\vspace*{0.3 cm}
\begin{tablenotes}
\normalsize
\item
Each row begins with the name of the data set. Yeast datasets are named by the
systematic names of the
two {\it S. cerevisiae} paralogous open reading frames.
The
``EDN/ECP'' row represents the primate EDN/ECP data set.  The  ``Introns''
row
represents the segmentally-duplicated primate introns.
The ``lnL IS-IGC''
column shows the maximum log-likelihood value of the HKY+IS-IGC model.
The ``Diff'' column specifies the number of log-likelihood units
by which the HKY+IS-IGC value exceeds the maximum log-likelihood value of
same model when $\tau$ is constrained to $0$. The ``$\tau$'' column
shows the estimated $\tau$ value from the HKY+IS-IGC model. The ``$r_2$'' and
``$r_3$'' columns show the estimated relative substitution rates of
the second and third codon positions from the HKY+IS-IGC model.
The column labelled ``HKY Prop''
shows the estimated proportions of
nucleotide changes attributable to IGC with
the HKY+IS-IGC model.
The column labelled ``MG94 Prop'' shows the estimated
proportions of codon changes attributable
to IGC with
the MG94+IS-IGC model.
``N.A.'' is written in table entries to denote ``not applicable''.
\end{tablenotes}
\end{table}

\newpage
\begin{table}[h]
\caption{IGC parameter estimates.}
\centering
\begin{tabular}{l ccccc ccccc}
\toprule
 & \multicolumn{2}{c}{{\bf \shortstack{$\tau$ \\ IS-IGC}}} & {\bf \shortstack{$\tau$ \\ PS-IGC} } & \multicolumn{2}{c}{\bf \shortstack{$1/p$ \\ PS-IGC} }  \\
\cmidrule(lr){2-3} \cmidrule(lr){4-4} \cmidrule(lr){5-6}
{\bf Paralog Pair}     & MCLE   & Interquartile & MCLE & MCLE & Interquartile   \\
\midrule
YLR406C,YDL075W & 5.10 & (4.37, 6.12) & 5.10 & 4.04 & (2.97, 4.87)\\
YER131W,YGL189C & 5.27 & (3.89, 6.75) & 5.27  & 13.24 & (9.46, 16.91)  \\
YML026C,YDR450W & 12.85 & (10.46, 15.32) & 12.84 & 1.04 & (1.00, 1.27)\\
YNL301C,YOL120C & 7.94 & (6.88, 8.71) & 7.93 & 103.52 & (68.27, 141.12) \\
YNL069C,YIL133C & 3.63 & (3.15, 4.11) & 3.63 & 12.37 & (8.90, 13.20)\\
YMR143W,YDL083C & 9.19 & (7.28, 13.85) & 9.20 & 4.79 & (3.55, 6.45) \\
YJL177W,YKL180W & 6.45 & (5.13, 7.20) & 6.45 & 8.50 & (7.11, 10.53)\\
YBR191W,YPL079W & 13.66 & (10.99, 15.41) & 13.65 & 9.08 & (7.57, 10.78)\\
YER074W,YIL069C & 20.90 & (16.69, 24.06) & 20.89 & 53.05 & (38.86, 72.06) \\
YDR418W,YEL054C & 5.16 & (4.23, 6.48) & 5.16 & 3.80  & (3.13, 4.69)\\
YBL087C,YER117W & 11.05 & (8.76, 13.22) & 11.03 & 29.45 & (20.14, 38.67) \\
YLR333C,YGR027C & 9.88 & (8.11, 11.43) & 9.85 & 36.68  & (26.96, 53.72) \\
YMR142C,YDL082W & 14.37 & (11.80, 16.77) & 14.36 & 33.61 & (29.45, 46.62) \\
YER102W,YBL072C & 14.77 & (12.25, 16.36) & 14.76 & 26.34 & (20.61, 61.41) \\
EDN/ECP        & 1.79  & (1.40, 2.21) & 1.79  & 6.90 & (4.32, 8.34)  \\
Introns        & 0.44  & (0.41, 0.46) & 0.52 & 370.4 & (142.9, 523.0) \\
\bottomrule
\label{tab:IGCestimates}
\end{tabular}
\vspace*{0.3 cm}
\begin{tablenotes}
\normalsize
\item
Each row begins with the name of the data set, as detailed in the caption
to Table \ref{tab:IS-IGCYeastResults}.
Columns labelled ``MCLE'' contain maximum composite likelihood estimates
and columns labelled ``Interquartile'' represent the $25{\mbox{th}}$ and $75{\mbox{th}}$ percentiles of the estimated values in $100$ parametric bootstrap samples.
The ``$\tau$, IS-IGC'' columns show the $\tau$ values
estimated with
the HKY+IS-IGC model.
The ``$\tau$, PS-IGC'' column shows the $\tau$ values
estimated with
the HKY+PS-IGC model.  Because the parametric bootstrap analyses for
the HKY+PS-IGC model did not consider uncertainty in the $\tau$ parameter (see Materials
and Methods), interquartile ranges are not available for these $\tau$ estimates.
The ``$1/p$, HKY+PS-IGC'' columns show
the average tract length in nucleotides of fixed IGC events as estimated
with
the HKY+PS-IGC model.
\end{tablenotes}
\end{table}

\end{document}